\documentclass[osa,twocolumn,showpacs,floatfix]{revtex4}
\usepackage{graphicx}%
\usepackage{dcolumn}
\usepackage{amsmath}
\begin{document}

\title{Spatial Solitons in Optically-Induced Gratings}
\author{Dragomir Neshev, Elena Ostrovskaya, and Yuri Kivshar}
\affiliation{Nonlinear Physics Group, Research School of Physical
Sciences and Engineering, Australian National University,
Canberra ACT 0200, Australia}
\author{Wieslaw Krolikowski}
\affiliation{Laser Physics Center, Research School of Physical Sciences and Engineering, Australian National University, Canberra
ACT 0200, Australia}

\begin{abstract}
We study experimentally nonlinear localization effects in
optically-induced gratings created by interfering plane waves in a
photorefractive crystal. We demonstrate the generation of spatial
bright solitons similar to those observed in arrays of coupled
optical waveguides. We also create pairs of out-of-phase solitons
which resemble ``twisted'' localized states in nonlinear lattices.
\end{abstract}

\maketitle
Formation of spatial optical solitons in periodic refraction index 
gratings formed by arrays of coupled optical waveguides was first 
studied theoretically by Christodoulides and Joseph~\cite{chrjos88}.
A standard theoretical approach to study the nonlinear localization 
of light  in gratings is based on the decomposition of the total
electric field in a sum of {\em weakly coupled} fundamental modes
excited in each waveguide of the array; a similar approach is
known in solid-state physics as {\em the tight-binding
approximation}. Then, the wave dynamics can be described by an
effective discrete nonlinear Schr\"odinger (NLS)
equation~\cite{chrjos88,kiv93,ace_et96,led_et01}, that has stationary, 
spatially localized solutions in the form of {\em discrete optical 
solitons}. Discrete solitons have
been extensively explored in a number of theoretical papers (see,
e.g., Refs.~\cite{kiv93,led_et01}), and also generated
experimentally~\cite{eis_et98,silste01}.  On the other
hand, several recent studies \cite{sukh_pre02, efr_et02,ost_et02}
indicate that the dynamics of nonlinear localized modes in
periodic structures cannot always be adequately described within
the discrete (tight-binding) approximation, and the full study of the
continuous models with periodic potentials is more accurate.

Recently, Fleischer {\it et al.} \cite{fle_et02} reported the first
experimental observation of spatial optical solitons in an array
of {\em optically-induced} waveguides. They created optical
gratings by interfering pairs of plane waves in a photorefractive
crystal, and employed screening nonlinearity to observe
self-trapping of a probe beam in the form of staggered and
unstaggered localized modes. Being inspired by these observations
\cite{fle_et02}, in this Letter we study experimentally novel
self-trapping effects in optically-induced gratings. We create a
well-controllable index grating by plane-wave interference in a
photorefractive crystal, and then demonstrate the
excitation of odd and even nonlinear localized states and their
bound states.

Following Efremidis {\em et al.} \cite{efr_et02}, we consider a
photorefractive Strontium Barium Niobate (SBN:60) crystal as a
nonlinear medium. The SBN sample is assumed to be externally
biased with a DC field along its extraordinary $x$-axis
(crystalline $c$-axis). Optically-induced refractive index
gratings  are created in the crystal via photorefractive effect,
by periodic intensity patterns  resulting from the interference of
two or more ordinary-polarized plane waves (wide beams of high
intensity, polarized along the ordinary $y$-axis). Then we study
evolution of the extraordinary polarized probe beam propagating
through this periodic structure \cite{efr_et02}. Since the
relevant electro-optic coefficients of the SBN crystal are
substantially different for orthogonal polarizations, the grating
created by ordinary polarized beams can be treated as being
essentially one-dimensional and constant along propagation 
$z$-axis. This one-dimensional grating, created by the interference of
plane waves with wavelength $\lambda$ has the intensity pattern
$I_g = I_0 \cos^2[K x]$, where $K=n_0 k_0 \sin \theta$, $\theta$
is the angle between a plane wave and the $z$-axis, $k_0 =
2\pi/\lambda$, and $n_0$ is the refractive index along the
ordinary axis.

Provided that the intensity of the probe beam, $|E|^2$, is weaker
than that of the grating, $I_0$, one can assume that the
back-action of the the probe beam on the grating is negligible.
Then the evolution of a probe beam in the grating is governed by
the following equation \cite{efr_et02}
\begin{equation}
\label{NLS}
2 i k_1 \frac{\partial E}{\partial z} +
\frac{\partial^2 E}{\partial x^2} - k_0^2 n_e^4 r_{33}
{\cal E}_{\rm sc} E =0,
\end{equation}
where $n_e$ is the refractive index along the extraordinary axis,
$k_1=k_0n_e$, and ${\cal E}_{\rm sc}$ is the space-charge field,
${\cal E}_{\rm sc} = {\cal E}_0/(1+I)$, expressed through the
total intensity, $I = |E|^2 + I_g$, normalized with respect to the
dark irradiance of the crystal, $I_d$. For our choice of the
polarity of the biasing DC field, $\gamma_{\rm nl} \equiv k_0^2 n_e^4
r_{33} {\cal E}_0 >0$, and the nonlinearity exhibited by the probe beam is
{\em self-focusing}.

\begin{figure}
\centerline{\includegraphics[width=2.8in]{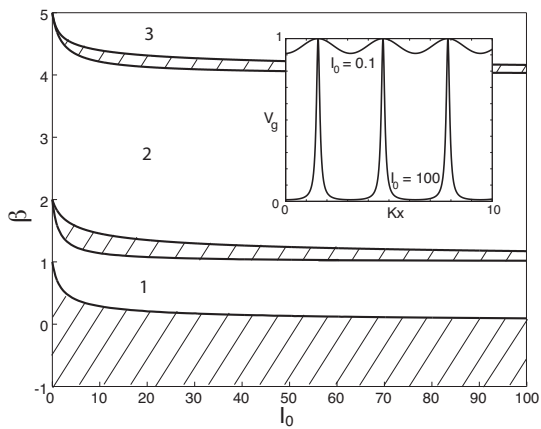}} 
\vskip -0.1in 
\caption{Band structure of the energy spectrum, $\beta$, in
the linear grating potential $V^{lin}_g=[1+I_0 \cos^2(Kx)]^{-1}$ 
(for $\gamma_{\rm nl}=1$):
numbered - Bloch bands, hashed - band gaps. Inset: effective
potential $V^{\rm lin}_g$ shown for two values of the 
grating intensity ($I_0$).}
\label{discr_fig1} \vskip -0.1in
\end{figure}

When $|E|^2 \ll I_0$, the probe beam experiences effectively
linear grating. This grating is {\em shallow} when $I_0<1$, and
{\em deep} when $I_0 \gg 1$. In this case, Eq.~(\ref{NLS})
can be presented as the linear Schr\"odinger equation with a
periodic potential, 
$V^{\rm lin}_g\approx \gamma_{\rm nl}/ [1+I_0 \cos^2(K x)]$, with
its stationary modes found in the form $E(z,x)=E_0(x)\exp[i
(\beta/k_1) z]$. The energy spectrum, $\beta$, of the waves guided
by the grating, as a function of $I_0$, displays a band-gap
structure (see Fig.~\ref{discr_fig1}). The spectral bands
correspond to the regime when the beam spreads across the entire
lattice; these extended states are Bloch waves of the periodic
structure. In the gaps separating the bands, no periodic guided
modes exist. Even when the effective linear grating becomes very
deep, the Bloch-wave bands remain wide, which means that the
coupling between the individual waveguides in the grating remains
very strong. As a result, the tight-binding approximation and the
corresponding discrete NLS model, valid for a weak coupling, may
be not valid for optical gratings induced in nonlinear saturable
media.

\begin{figure}
\centerline{\includegraphics[width=3.1in]{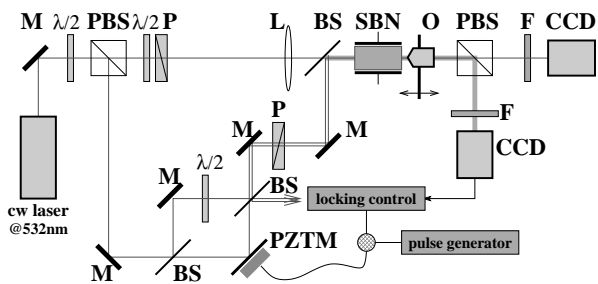}}
\caption{Experimental setup: M - mirrors; PZTM - piezo
translational mirror; PBS - polarizing beam splitters; BS - plane
beam splitters; P - polarizer; SBN - photorefractive crystal; O - 
microscope objective; $\lambda/2$ - half-wave plate; L - AR-coated 
lens; F - filters; CCD - cameras.} 
\label{discr_setup}
\vskip -0.1in
\end{figure}

When the probe-beam intensity grows, the effective grating becomes
{\em nonlinear}, as the beam experiences self-focusing. 
In this case, the probe beam can be trapped in the form of
spatially localized modes - bright spatial solitons. For the focusing 
nonlinearity, these
localized modes are found in the semi-infinite gap (bellow 
band $1$ in Fig.~\ref{discr_fig1}), and they resemble conventional
bright solitons modulated by the lattice. These modes are analogous 
to {\em unstaggered discrete solitons} of the nonlinear lattices
only in the tight-binding regime which has a limited applicability
in our case. However, this analogy can still be employed to
understand the structure of the localized states, even though the
saturable nature of nonlinearity does not allow us to apply the
discrete model directly.

\begin{figure}
\centerline{\includegraphics[width=3.1in]{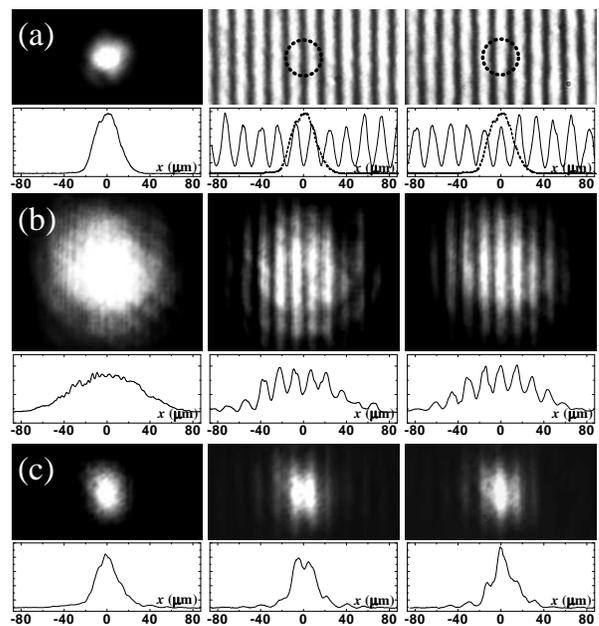}} 
\vskip -0.1in 
\caption{Experimental demonstration of odd and even spatial
solitons. (a) input beam and optical lattice (power 23$\mu$W); (b)
the output probe beam at low power ($2\times10^{-3}\mu$W); (c)
localized states ($87\times10^{-3}\mu$W). Left column - vibrating
PZTM, middle - even excitation, right - odd excitation. Electric 
field is 3600V/cm.} 
\label{discr_solitons} 
\vskip -0.1in
\end{figure}

Recently, the model similar to Eq.~(\ref{NLS}) has been analyzed,
beyond the tight-binding approach, in the context of coherent
matter waves in optical lattices \cite{ost_et02}. It was shown
that it supports a number of nonlinear localized states, some of
which are analogous to the solitons of the discrete NLS equation.
Moreover, the localized modes were found to exist in the form of
multi-soliton bound states, and one of the simplest bound states
is a two-soliton state that corresponds to the so-called ``twisted
mode'' of discrete lattices \cite{led_et01,sukh_pre02}. Below, we 
demonstrate experimentally
the existence of different types of nonlinear localized states in
an optically-induced grating in the regime of the self-focusing
nonlinearity.

\begin{figure}
\centerline{\includegraphics[width=2.5in]{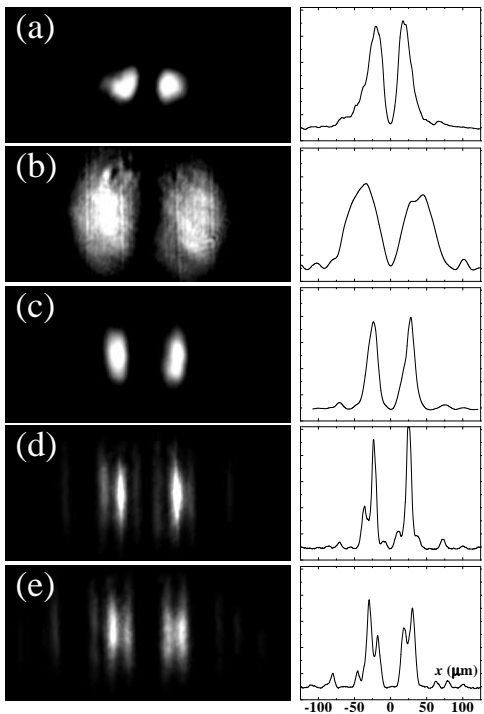}} 
\vskip -0.1in 
\caption{Experimental demonstration of the soliton bound
states--twisted modes: (a) input; (b-e) outputs: (b) linear
diffraction; (c) dipole beam; (d) even symmetry twisted mode; (e)
odd symmetry twisted mode. Powers: grating - 23$\mu$W, dipole
beam - 0.12$\mu$W. Intensity ratio is 1.3, electric field is 3600V/cm.}
\label{twisted_modes} 
\vskip -0.15in
\end{figure}

The experimental setup is shown in Fig.~\ref{discr_setup}. The
light of a 
Nd:YVO$_4$ laser at 532nm is split into two parts.
The transmitted (extra-ordinary polarized) beam 
is focused by a 6-cm (or 5-cm) focal-length lens on the input face of
a 15mm long SBN:60 crystal. The second,
ordinary-polarized beam is used to form the optical lattice. It
passes through a Mach-Zehnder interferometer aligned such that
its two output beams 
intersect at a small angle, thus
producing interference fringes inside the crystal. The difference
in the optical paths of the interferometer could be locked by use 
of a piezo-controlled mirror, in order to stabilize the 
interference pattern. It also enabled for precise positioning of 
the interference maxima in respect to the location of the probe 
beam, which propagates parallel to the induced waveguides. 
The input and output faces of the crystal could be imaged by a
microscope objective onto the two CCD cameras. Additional
white-light illumination is used to vary the ``dark'' irradiance
of the crystal and subsequently control degree of saturation of
the nonlinearity.


Similar to the localized modes of discrete nonlinear lattices,
spatial solitons in optically-induced gratings can be ``even'',
i.e. centered between two induced waveguides (minimum of the
grating intensity), or ``odd'', i.e. centered on the waveguide
(maximum of the grating intensity). We observed both types of
the localized modes.
At low intensities of the probe beam 
($I_0 /|E|^2 =50$) we observed diffraction resembling the formation 
of Bloch waves of the periodic potential [Fig.~\ref{discr_solitons}]. 
Diffraction pattern is almost independent on the initial position 
of the probe beam relative to the grating [(b)]. However, when the 
intensity of the probe beam is increased ($I_0 /|E|^2 =1.2$) two 
distinct states form [(c)]. When the maximum of the probe beam is 
centered between the maxima of the grating, an even state is 
generated [(c) - middle column]. However, when the probe beam is 
centered on a maximum,
an odd state forms [(c) - right column]. The even 
mode was observed to be unstable, and transforms into a 
non-symmetric structure due to small perturbations of the beam 
position.
The qualitative picture described above does not change if a smaller 
input beam is used (focusing with a 5-cm lens), such that only a
single waveguide is initially excited. If the PZT mirror vibrates 
with high frequency, the grating disappears, and a standard
single soliton state is observed [(c) - left column].


The simplest ``twisted'' mode is similar to a pair of two
out-of-phase solitons; it can also be odd or even. The twisted
modes are generated by introducing a tilted glass plate in half of
the input beam. The tilt is set such that both parts of the beam
are $\pi$-phase shifted, Fig.~\ref{twisted_modes}. The input beam
[(a)] can be centered in between two maxima or on a maximum of the
optical lattice. Without a DC field, the dipole beam
diffracts [(b)]. When a voltage is applied and the PZT mirror is
set to vibrate, a pair of two repelling bright solitons is
observed [(c)]. When the optical lattice is formed, a pair of
out-of-phase ``odd'' states is created [(d)]. If the beam is
centered on a maximum of the optical lattice, a pair of ``even''
states is observed with a central waveguide not being excited
at all [(e)].


In conclusion, we have demonstrated experimentally the generation
of spatial optical solitons in optically-induced gratings created
by interfering plane waves in a photorefractive crystal. Since the
light-induced optical lattices provide 
much greater control of the
grating parameters than fabricated waveguide arrays, we believe
that these results open many new possibilities for the study of
various nonlinear effects in optically induced gratings, including
gap solitons.

The work has been partially supported by the Australian
Photonics Cooperative Research Center and the Australian Research
Council. We thank Glen McCarthy and Andrey Sukhorukov for useful
discussions.

\vspace{-0.5 true cm}

\end{document}